\documentclass[10pt,prb,aps,superscriptaddress,showpacs,twocolumn]{revtex4}
\usepackage{amssymb,amsmath}
\usepackage{graphicx}
\usepackage{color}
\usepackage{natbib}
\usepackage{hyperref}
\usepackage{setspace}
\usepackage{floatrow}
\usepackage[english]{babel}
\usepackage{sidecap} 
\usepackage[T2A,OT1]{fontenc}
\usepackage[utf8]{inputenc}

\begin{document}

\newcommand\cyrtext[1]{{\fontencoding{T2A}\selectfont #1}}  
\newcommand{\df}[2]{\frac{\partial #1}{\partial #2}}             
\newcommand{\ds}[2]{\frac{{\partial}^2 #1}{\partial {#2}^2}}     
\newcommand{\tens}[1]{\hat{#1}}
\newcommand{\bas}[1]{\vec{e}_{#1}} 
\newcommand{\Gk}[2]{G_{\vec{k}}^{#1#2}} 
\newcommand{\refPR}[1]{[\onlinecite{#1}]}
\newcommand{\correction}[1]{{\color{red} #1}}
\newcommand{\av}[1]{\left< #1 \right>} 

\title{Nonlocal homogenization for nonlinear metamaterials}
\author{Maxim~A.~Gorlach}
\email{Maxim.Gorlach.blr@gmail.com}
\affiliation{ITMO University, St. Petersburg 197101, Russia}
\author{Tatiana A. Voytova}
\affiliation{ITMO University, St. Petersburg 197101, Russia}
\author{Mikhail Lapine}
\affiliation{School of Mathematical and Physical Sciences, University of Technology Sydney, NSW 2007, Australia}
\author{Yuri~S.~Kivshar}
\affiliation{Nonlinear Physics Centre, Australian National University, Canberra, ACT 0200, Australia}
\affiliation{ITMO University, St. Petersburg 197101, Russia}
\author{Pavel~A.~Belov}
\affiliation{ITMO University, St. Petersburg 197101, Russia}

\begin{abstract}
We present a consistent theoretical approach for calculating effective nonlinear susceptibilities of metamaterials taking into account  both frequency and spatial dispersion. Employing the discrete dipole model, we demonstrate that effects of spatial dispersion become especially pronounced in the vicinity of effective permittivity resonance where nonlinear susceptibilities reach their maxima. In that case spatial dispersion may enable simultaneous generation of two harmonic signals with the same frequency and polarization but different wave vectors. We also prove that the derived expressions for nonlinear susceptibilities transform into the known form  when spatial dispersion effects are negligible. In addition to revealing new physical phenomena, our results provide useful theoretical tools for analysing resonant nonlinear metamaterials.
\end{abstract}

\pacs{42.65.Pc, 42.70.Mp, 42.65.An}
\maketitle
%

\section{Introduction}\label{sec:Introduction}
The field of nonlinear metamaterials attracts significant research interest due to the numerous fascinating applications~\refPR{Lapine-2014,Shadrivov-book,Zheludev,Minovich}. The use of large nonlinearities available in resonant nonlinear metamaterials~\refPR{Lapine-2014} opens a possibility to all-optical signal processing~\refPR{Soljacic} and provides reach opportunities to implement tunable and reconfigurable photonic devices~\refPR{Lapine-2011,Slob}.

A fundamental task existing in the field is characterization of electromagnetic properties of nonlinear metamaterials, i.e. homogenization. For many nonlinear scenarios nonlinearity of metamaterials is described in a perturbative way in terms of nonlinear susceptibilities~\refPR{Boyd}. This allows one to exploit the framework of nonlinear optics and directly compare nonlinearities of artificially structured media with those occuring in natural crystals. To date, a number of approaches to homogenize nonlinear composites and metamaterials were  reported~\refPR{Sipe,Lapine-2003,Giordano,Rose2012,Silv2013}. However, most of these approaches do not take spatial dispersion into account. Spatial dispersion (or nonlocality) implies the dependence of polarization of a physically small volume on the fields existing in the neighboring regions of space. A number of theoretical and experimental studies demonstrate that nonlocality is pronounced in a wide class of linear metamaterials~\refPR{Simovski2009,Orlov,Zhukovsky,Gorlach2015}. In linear structures  nonlocality is often described in terms of effective permittivity tensor depending on both frequency and wave vector. Comprehensive theoretical models describing nonlocality in linear artificial composites were developed~\refPR{Silv2007,Silv2007L,Alu} and, in particular, a generalization of the Clausius-Mossotti formula for the case of discrete three-dimensional (3D) metamaterial was proposed~\refPR{Silv2007L}. However, to the best of our knowledge, the consistent theoretical description of spatial dispersion effects in nonlinear metamaterials remains an open problem so far.

    \begin{figure}[b]
    \includegraphics[width=0.7\linewidth]{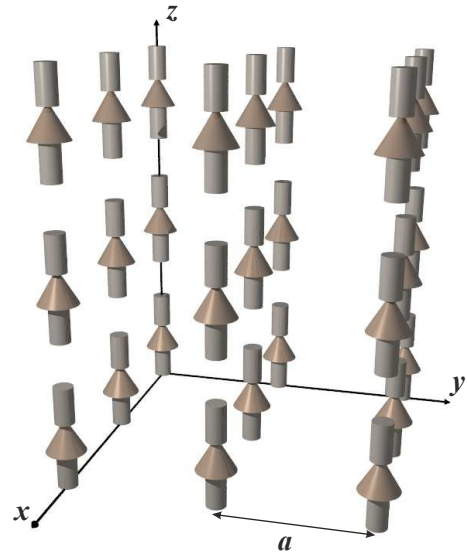}
    \caption{(Color online) Schematic representation of a three-dimensional metamaterial composed of nonlinear uniaxial inclusions.}
    \label{ris:Uniaxial}
    \end{figure}

In this paper, we derive nonlocal nonlinear susceptibilities of a 3D nonlinear metamaterial composed of uniaxial scatterers~(Fig.~\ref{ris:Uniaxial}). We employ the discrete dipole model~\refPR{Purcell,Draine,Belov2005} describing the field of the  scatterer in dipole approximation, whereas the properties of the individual meta-atom are characterized by linear and nonlinear  polarizability tensors. The rest of the paper is organized as follows. In Sec.~\ref{sec:Homogenization} we derive general expressions for nonlocal nonlinear susceptibilities of discrete metamaterial. Sec.~\ref{sec:Local} demonstrates that in the limiting case when spatial dispersion effects in the structure can be neglected our results reduce to the known expressions for local field corrections. In Sec.~\ref{sec:Numerical} we illustrate the main features of the developed approach providing a numerical example for the structure composed of short wires loaded with varactor diodes. In particular, we highlight the essential role of spatial dispersion effects at frequencies in the vicinity of effective pemittivity resonance. Finally, in Sec.~\ref{sec:Discussion} we discuss the obtained results. Calculation of linear and nonlinear polarizabilities of a short varactor-loaded wire is provided in Appendix.

\section{Nonlocal homogenization of discrete nonlinear metamaterial}\label{sec:Homogenization}
We consider an array of nonlinear uniaxial meta-atoms arranged in a cubic 3D lattice with the period $a$ (Fig.~\ref{ris:Uniaxial}). Note that chaotic arrangement of the similar nonlinear dipoles was studied in Ref.~\refPR{Kalinin}. However, in that work mutual interactions of meta-atoms were not considered.  In the present derivation we use the CGS system of units and assume $e^{-i\omega t}$ time dependence of monochromatic fields. We consider excitation of nonlinear structure by an incident wave with frequency $\omega$, denoting wave vector of the fundamental wave propagating in the metamaterial by $\vec{k}$. Due to the nonlinear nature of inclusions, the incident wave generates polarization not only at the fundamental frequency $\omega$, but also at frequencies $2\,\omega$, $3\,\omega$, etc. This nonlinear polarization becomes a source of harmonics at frequencies $2\,\omega$, $3\,\omega$, etc. In the present analysis we consider only second and third harmonics omitting nonlinear contributions of higher order. This is a typical assumption for many nonlinear metamaterials~\refPR{Lapine-2014}. Furthermore, for the sake of simplicity we assume that the scatterers can be polarized only along $z$ axis. Then the only essential components of linear and nonlinear susceptibility tensors would be $\chi^{(1)}_{zz}$, $\chi^{(2)}_{zzz}$ and $\chi^{(3)}_{zzzz}$. Accordingly, the subscript $z$ is omitted further in the designations of vector and tensor components. Under these assumptions the dipole moment $d$ of the individual meta-atom is given by the equations:
\begin{align}
& d(\omega)=\alpha_1(\omega)\,E_{\rm{tot}}(\omega)+\notag\\
& 2\,\alpha_2(\omega;2\,\omega,-\omega)\,E_{\rm{tot}}(2\,\omega)E_{\rm{tot}}^*(\omega)+\label{d1}\\
& 3\,\alpha_3(\omega;\omega,\omega,-\omega)\,\left|E_{\rm{tot}}(\omega)\right|^2\,E_{\rm{tot}}(\omega)\:,\notag\\
\notag \\
& d(2\,\omega)=\alpha_1(2\,\omega)\,E_{\rm{tot}}(2\,\omega)+\notag\\
& \alpha_2(2\,\omega;\omega,\omega)\,E_{\rm{tot}}^2(\omega)\:,\label{d2}\\
\notag \\
& d(3\,\omega)=\alpha_1(3\,\omega)\,E_{\rm{tot}}(3\,\omega)+\notag\\
& 2\,\alpha_2(3\,\omega;2\,\omega,\omega)\,E_{\rm{tot}}(2\,\omega)\,E_{\rm{tot}}(\omega)+\label{d3}\\
& \alpha_3(3\,\omega;\omega,\omega,\omega)\,E_{\rm{tot}}^3(\omega)\:.\notag
\end{align}
where $\alpha_1$, $\alpha_2$ and $\alpha_3$ stand for linear, second- and third-order nonlinear polarizabilities. The total field $E_{\rm{tot}}(\omega)$ is the sum of the external field acting on the scatterer $E(\omega)$ and field associated with radiation friction~\refPR{landau2} $E_s(\omega)=2i\omega^3\,d(\omega)/(3\,c^3)$. In this case polarizabilities introduced in Eqs.~\eqref{d1}, \eqref{d2}, \eqref{d3} are so-called {\it bare} polarizabilities, i.e.\ they do not contain the radiation loss contribution~\refPR{SipeK}. Note that bare polarizability of a lossless scatterer is purely real. An alternative  description of radiation losses incorporates an imaginary part into scatterer polarizability~\refPR{SipeK,Belov2005}. The latter approach, however, turns out to be less convenient for nonlinear structures. Linear polarizability $\alpha_1(\omega)$ along with nonlinear polarizabilities $\alpha_2$ and $\alpha_3$ can be calculated for a particular scatterer. For example, the calculation of nonlinear polarizabilities of varactor-loaded short wire is provided in Appendix, whereas the analysis of nonlinear properties of varactor-loaded split-ring resonators is carried out in Refs.~\refPR{Lapine-2004,Powell-2007,Poutrina}. It should be emphasized that the field $E(\omega)\equiv E_{\rm{tot}}(\omega)-E_s(\omega)$ appearing in Eqs.~\eqref{d1}, \eqref{d2}, \eqref{d3} is {\it local field}, i.e.\ the electric field in the point where the scatterer is located. On the other hand, in the definition of effective material parameters the {\it average field} appears. The average fields are defined as~\refPR{Silv2007,Alu}
\begin{gather}
\left<\vec{E}(\Omega)\right>=\frac{1}{V_0}\,\int\limits_{V_0}\,\vec{E}(\Omega;\vec{r})\,e^{-i\vec{K}\cdot\vec{r}}\,dV\:,\label{Eav}\\
\left<\vec{P}(\Omega)\right>=\frac{1}{V_0}\,\int\limits_{V_0}\,\vec{P}(\Omega;\vec{r})\,e^{-i\vec{K}\cdot\vec{r}}\,dV\:,\label{Pav}
\end{gather}
where $V_0=a^3$ is a unit cell volume, $\Omega$ is arbitrary frequency and vector function $\vec{K}=\vec{K}(\Omega)$ is specified later in this section. The average structure polarization is
\begin{equation}\label{AvP}
\left<P(\Omega)\right>=d(\Omega)/V_0\:,
\end{equation}
and the average electric field can be related to the average polarization by \refPR{Alu}
\begin{equation}\label{AvField}
\left<E(\Omega)\right>=\Gamma_{k}(\Omega,\vec{K})\,d(\Omega)\:,
\end{equation}
with
\begin{equation}\label{Gamma}
\Gamma_k(\Omega;\vec{K})=-\frac{4\pi}{a^3}\,\frac{\Omega^2-K_z^2\,c^2}{\Omega^2-K^2\,c^2}\:.
\end{equation}
Local field acting on the scatterer in the coordinate origin can be evaluated via dyadic Green's function~\refPR{Novotny} $\widehat{G}(\Omega;\vec{r})$ and dipole moments $d_{mnl}$ of metaatoms as
\begin{equation}
E(\Omega)=\sum\limits_{(m,n,l)\not=(0,0,0)}\,G_{zz}(\Omega;-\vec{r}_{mnl})\,d_{mnl}(\Omega)
\end{equation}
where the indices $m,n,l$ enumerate the lattice sites and the dipole moments of the scatterers in the unbounded structure are distributed as
\begin{equation}
d_{mnl}(\Omega)=d(\Omega)\,e^{i\vec{K}(\Omega)\cdot\vec{r}_{mnl}}\:.
\end{equation}
The distribution of polarization at the fundamental frequency is determined by the incident wave, and in this case $\vec{K}(\omega)=\vec{k}$, where $\vec{k}$ is a wave vector of the structure eigenmode. Since second-order nonlinear polarization is a quadratic function of the fundamental wave, $\vec{K}(2\,\omega)=2\,\vec{k}$ and similarly $\vec{K}(3\,\omega)=3\,\vec{k}$. Thus, the expression for local field can be represented as
\begin{equation}\label{LocField}
E(\Omega)=G_k\left(\Omega;\vec{K}(\Omega)\right)\,d(\Omega)\:,
\end{equation}
with the lattice sum defined as

\begin{equation}
G_k(\Omega;\vec{K})\equiv\sum\limits_{(m,n,l)\not=(0,0,0)}\,G_{zz}(\Omega;\vec{r})\,e^{-i\vec{K}\cdot\vec{r}_{mnl}}\:.
\end{equation}
Therefore, the total field appearing in Eqs.~\eqref{d1}, \eqref{d2}, \eqref{d3} is
\begin{equation}\label{TotField}
E_{\rm{tot}}(\Omega)=G_k'\left(\Omega,\vec{K}(\Omega)\right)\,d(\Omega)
\end{equation}
with $G_k'(\Omega,\vec{K})=G_k(\Omega,\vec{K})+2i\,\Omega^3/(3\,c^3)$. Effective algorithms for the lattice sum rapid evaluation where developed earlier~\refPR{Belov2005}. Importantly, it can be proved that for real $\Omega$ and $\vec{K}$ ${\rm Im}\, G_k'(\Omega,\vec{K})=0$~\refPR{Belov2005}. Making use of Eqs.~\eqref{TotField} and \eqref{d1} it is easy to see that the dispersion equation for the linear structure with the metaatom's polarizability $\alpha_1(\omega)$ has the form~\refPR{Coevorden,Belov2005,Gorlach2014}
\begin{equation}\label{DispEq}
\alpha_1^{-1}(\omega)-G_k'(\omega;\vec{k})=0\:.
\end{equation}
Using Eqs.~\eqref{AvField} and \eqref{LocField} we obtain
\begin{equation}\label{TotAvField}
E_{\rm{tot}}(\Omega)=\Phi(\Omega;\vec{K})\,\left<E(\Omega)\right>
\end{equation}
where $\Phi(\Omega;\vec{K})=G_k'(\Omega;\vec{K})/\Gamma_k(\Omega;\vec{K})$ is introduced for convenience. Making use of Eqs.~\eqref{TotField} and \eqref{AvField}, the relation between local and averaged fields can be also represented in the alternative form
\begin{equation}\label{TotAvField2}
E_{\rm{tot}}(\Omega)=\left<E(\Omega)\right>+C_k(\Omega;\vec{K})\,d(\Omega)
\end{equation}
where $C_k(\Omega;\vec{K})=G_k'(\Omega;\vec{K})-\Gamma_k(\Omega;\vec{K})$ is the lattice interaction constant. Finally, inserting the expressions Eqs.~\eqref{AvP}, \eqref{TotAvField}, \eqref{TotAvField2} into Eqs.~\eqref{d1}, \eqref{d2}, \eqref{d3}, we obtain the relation between the averaged structure polarization and the averaged field:
\begin{align}
& \left<P(\omega)\right>=\chi^{(1)}(\omega,\vec{k})\,\left<E(\omega)\right>+\notag\\
& 2\,\chi^{(2)}(\omega;2\,\omega,-\omega,\vec{k})\,\left<E(2\,\omega)\right>\,\left<E(\omega)\right>^*+\label{P1}\\
& 3\,\chi^{(3)}(\omega;\omega,\omega,-\omega,\vec{k})\,\left|\left<E(\omega)\right>\right|^2\,\left<E(\omega)\right>\:,\notag\\
\notag\\
& \av{P(2\,\omega)}=\chi^{(1)}(2\,\omega,2\,\vec{k})\,\av{E(2\,\omega)}+\notag\\
& \chi^{(2)}(2\,\omega;\omega,\omega,2\,\vec{k})\,\av{E(\omega)}^2\:,\label{P2}\\
\notag\\
& \av{P(3\,\omega)}=\chi^{(1)}(3\,\omega,3\,\vec{k})\,\av{E(3\,\omega)}+\notag\\
& 2\,\chi^{(2)}(3\,\omega;2\,\omega,\omega,3\,\vec{k})\,\av{E(2\,\omega)}\,\av{E(\omega)}+\label{P3}\\
& \chi^{(3)}(3\,\omega;\omega,\omega,\omega,3\,\vec{k})\,\av{E(\omega)}^3\:.\notag
\end{align}
Nonlocal nonlinear susceptibilities in Eqs.~\eqref{P1}, \eqref{P2}, \eqref{P3} can be written in a compact form as follows:
\begin{widetext}
\begin{gather}
\chi^{(1)}(\Omega,\vec{K})=\frac{1}{a^3}\,\left[\alpha_1^{-1}(\Omega)-C_k(\Omega,\vec{K})\right]^{-1}\:,\label{Chi1}\\
\chi^{(2)}(\omega_3;\omega_2,\omega_1,\vec{K}(\omega_3))=\frac{\alpha_2(\omega_3;\omega_2,\omega_1)}{a^3\,\alpha_1(\omega_3)}\,\Phi(\omega_2,\vec{K}(\omega_2))\,\Phi(\omega_1,\vec{K}(\omega_1))\,\left[\alpha_1^{-1}(\omega_3)-C_k(\omega_3,\vec{K}(\omega_3))\right]^{-1}\:,\label{Chi2}\\
\chi^{(3)}(\omega_4;\omega_3,\omega_2,\omega_1,\vec{K}(\omega_4))=\frac{\alpha_3(\omega_4;\omega_3,\omega_2,\omega_1)}{a^3\,\alpha_1(\omega_4)}\Phi(\omega_3,\vec{K}(\omega_3))\,\Phi(\omega_2,\vec{K}(\omega_2))\,\Phi(\omega_1,\vec{K}(\omega_1))\,\left[\alpha_1^{-1}(\omega_4)-C_k(\omega_4,\vec{K}(\omega_4))\right]^{-1}\:.\label{Chi3}
\end{gather}
\end{widetext}
In Eq.~\eqref{Chi1}, $\Omega=\omega$, $2\,\omega$ or $3\,\omega$ and $\vec{K}(\Omega)=\vec{k}$, $2\,\vec{k}$ or $3\,\vec{k}$, respectively. In Eq.~\eqref{Chi2}, $\omega_3=\omega_2+\omega_1$, the pair $(\omega_2,\omega_1)$ acquires the values $(\omega,\omega)$, $(2\,\omega,\omega)$ and $(2\,\omega,-\omega)$. In Eq.~\eqref{Chi3}, $\omega_4=\omega_3+\omega_2+\omega_1$, the triplet $(\omega_3,\omega_2,\omega_1)$ acquires the values $(\omega,\omega,\omega)$ and $(\omega,\omega,-\omega)$. Note that Eqs.~\eqref{Chi1}, \eqref{Chi2} and \eqref{Chi3} are  valid for negative frequencies in which case $\Phi(-\Omega,\vec{K}(-\Omega))\equiv \Phi^*(\Omega,\vec{K}(\Omega))$.

Expressions \eqref{Chi1}, \eqref{Chi2} and \eqref{Chi3} depend implicitly on $\vec{k}$ which is the solution of the dispersion equation for linear structure Eq.~\eqref{DispEq}. Therefore, one may easily calculate nonlinear susceptibilities for a given direction of wave propagation and fundamental frequency $\omega$. Eqs.~\eqref{Chi1}, \eqref{Chi2}, \eqref{Chi3} suggest that nonlinear suscepbilities depend on the direction of wave propagation. Such effect is one of the manifestations of spatial dispersion. It is discussed in detail in Sec.~\ref{sec:Numerical}.

As additional evidence of the validity of our approach, we notice that the obtained expression Eq.~\eqref{Chi1} for the linear susceptibility of the structure coincides with the result derived in Ref.~\refPR{Silv2007L}. Furthermore, in the case of lossless scatterers and for the propagating mode with real $\omega$ and $\vec{k}$ effective nonlinear susceptibilities turn out to be purely real because the quantities $G_k'(\Omega,\vec{K})$ and $C_k(\Omega,\vec{K})$ are both real. This result satisfies energy conservation law.

\section{Comparison with the local effective medium model}\label{sec:Local}
In this section we demonstrate that in the limiting case when $K\,a\ll 1$ and $\Omega\,a/c\ll 1$, i.e.\ when  spatial dispersion effects in the structure are negligible, our results can be reduced to local nonlinear susceptibilities well-known from nonlinear optics. In this limit the interaction constant for a cubic lattice~\refPR{Belov2005,Silv2007L}
\begin{equation}\label{Cappr}
C_k(\Omega,\vec{K})=4\,\pi/(3\,a^3)
\end{equation}
for any real and sufficiently small $\Omega$ and $\vec{K}$. As a result, effective permittivity of the structure is given by the Clausius-Mossotti formula~\refPR{Collin}
\begin{equation}\label{Clausius}
\varepsilon_{\rm{loc}}(\omega)\equiv 1+4\,\pi\,\chi_{\rm{loc}}^{(1)}(\omega)=\frac{1+8\,\pi\,\alpha_1(\omega)/(3\,a^3)}{1-4\,\pi\,\alpha_1(\omega)/(3\,a^3)}\:.
\end{equation}
Equations~\eqref{Cappr} and \eqref{Clausius} yield also that
\begin{equation}\label{Prefactor}
\begin{split}
&\alpha_1^{-1}(\Omega)\,\left[\alpha_1^{-1}(\Omega)-C_k(\Omega,\vec{K}(\Omega))\right]^{-1}=\\
&\frac{1}{1-4\,\pi\,\alpha_1(\Omega)/(3\,a^3)}=\frac{\varepsilon_{\rm{loc}}(\Omega)+2}{3}\:.
\end{split}
\end{equation}
Furthermore, from Eq.~\eqref{Gamma} it is straightforward that
\begin{equation}\label{Gamma2}
\Gamma(\Omega,\vec{K}(\Omega))=\Gamma(\omega,\vec{k})\:,
\end{equation}
and as a consequence of Eqs.~\eqref{Cappr}, \eqref{Gamma2} and \eqref{DispEq}
\begin{equation}\label{GkTr}
\begin{split}
& G_k'(\Omega,\vec{K}(\Omega))\equiv\Gamma_k(\Omega,\vec{K}(\Omega))+C_k(\Omega,\vec{K}(\Omega))=\\
& \Gamma_k(\omega,\vec{k})+C_k(\omega,\vec{k})=G_k'(\omega,\vec{k})=\alpha_1^{-1}(\omega)\:.
\end{split}
\end{equation}
Expression for the factor $\Phi(\Omega,\vec{K}(\omega))$ can be transformed using Eqs.~\eqref{GkTr} and \eqref{Gamma2} as follows:
\begin{equation}\label{Phi}
\Phi(\Omega,\vec{K}(\Omega))\equiv\frac{G_k'(\Omega,\vec{K}(\Omega))}{\Gamma(\Omega,\vec{K}(\Omega))}=\frac{\alpha_1^{-1}(\omega)}{\alpha_1^{-1}(\omega)-4\pi/(3\,a^3)}\:.
\end{equation}
Thus, taking into account Eq.~\eqref{Clausius} we derive the simplified expression for the factor $\Phi$:
\begin{equation}\label{Phi2}
\Phi(\Omega,\vec{K}(\Omega))=\frac{\varepsilon_{\rm{loc}}(\omega)+2}{3}
\end{equation}
for any $\Omega=\omega,\,2\,\omega$ and $3\,\omega$. Applying the simplified expressions Eqs.~\eqref{Prefactor} and \eqref{Phi2} to the general formulas Eqs.~\eqref{Chi2} and \eqref{Chi3} we finally obtain:
\begin{widetext}
\begin{gather}
\chi^{(2)}_{\rm{loc}}(\omega_3;\omega_2,\omega_1)=\frac{\alpha_2(\omega_3;\omega_2,\omega_1)}{a^3}\,\frac{\varepsilon_{\rm{loc}}(\omega_3)+2}{3}\,\frac{\varepsilon_{\rm{loc}}(\omega_2)+2}{3}\,\frac{\varepsilon_{\rm{loc}}(\omega_1)+2}{3}\:,\label{Chi2Loc}\\
\chi^{(3)}_{\rm{loc}}(\omega_4;\omega_3,\omega_2,\omega_1)=\frac{\alpha_3(\omega_4;\omega_3,\omega_2,\omega_1)}{a^3}\,\frac{\varepsilon_{\rm{loc}}(\omega_4)+2}{3}\,\frac{\varepsilon_{\rm{loc}}(\omega_3)+2}{3}\,\frac{\varepsilon_{\rm{loc}}(\omega_2)+2}{3}\,\frac{\varepsilon_{\rm{loc}}(\omega_1)+2}{3}\:.\label{Chi3Loc}
\end{gather}
\end{widetext}
In Eq.~\eqref{Chi2Loc} $\omega_3=\omega_1+\omega_2$, in Eq.~\eqref{Chi3Loc} $\omega_4=\omega_3+\omega_2+\omega_1$. Both equations are also applicable for negative frequency values in accordance with $\varepsilon(-\Omega)\equiv \varepsilon^*(\Omega)$. Essentially, factors $\alpha_2/a^3$ and $\alpha_3/a^3$ describe second- and third-order nonlinear susceptibilities in the case when interaction of the scatterers is negligible. Factors $(\varepsilon_{\rm{loc}}+2)/3$ thus represent a local field correction to the nonlinear susceptibilities. The presented form of local field corrections is well-known in nonlinear optics~\refPR{Boyd}. Therefore, in the limit of negligible spatial dispersion our results are consistent with the previous studies based on the local effective medium approach.

\section{Numerical example}\label{sec:Numerical}

    \begin{figure}[h]
    \includegraphics[width=0.9\linewidth]{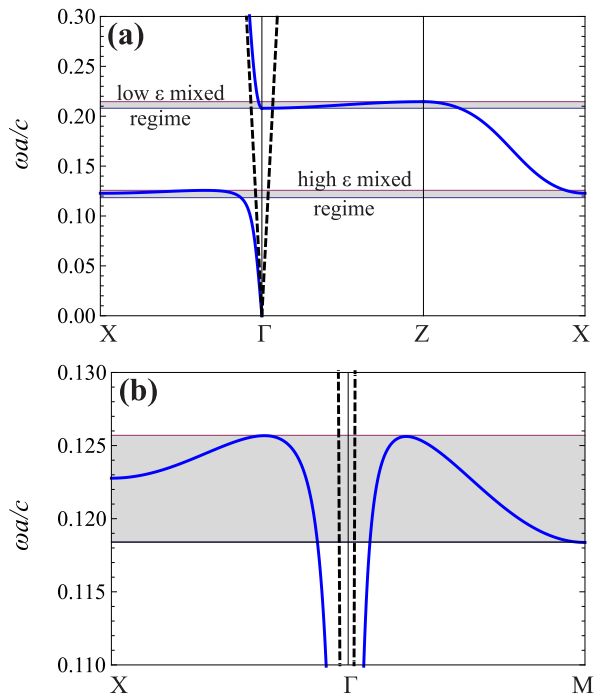}
    \caption{(Color online) (a) Calculated dispersion diagram for the structure composed of short varactor-loaded wires. (b) Enlarged branches $\Gamma$X and $\Gamma$M of the dispersion diagram in the vicinity of high-$\varepsilon$ mixed regime.}
    \label{ris:LinearDisp}
    \end{figure}

    \begin{figure}[t]
    \includegraphics[width=0.9\linewidth]{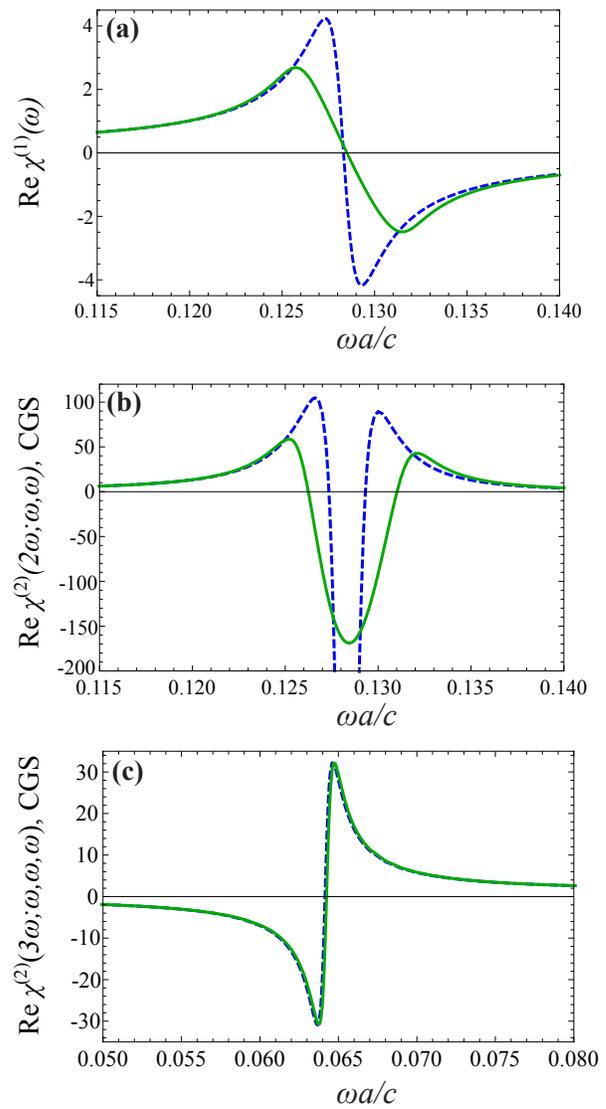}
    \caption{(Color online) Comparison of the developed theoretical approach (solid line) with the local effective medium model (dashed line). (a) Real part of metamaterial linear susceptibility in the vicinity of resonance $f_r$. (b, c) Real part of metamaterial second-order nonlinear susceptibility $\chi^{(2)}(2\omega;\omega,\omega)$ in the vicinity of the resonances (b) $f_r$ and (c) $f_r/2$.}
    \label{ris:Chi}
    \end{figure}

    \begin{figure}[t]
    \includegraphics[width=0.9\linewidth]{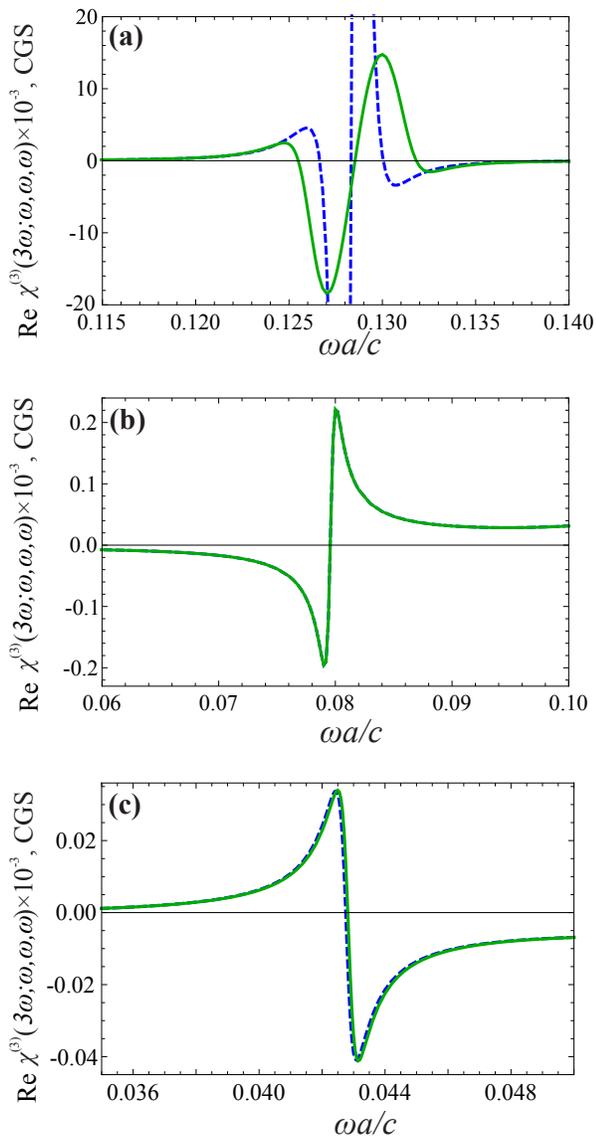}
    \caption{(Color online) Comparison of the developed theoretical approach (solid line) with the local effective medium model (dashed line). Real part of metamaterial third-order nonlinear susceptibility $\chi^{(3)}(3\,\omega;\omega,\omega,\omega)$ in the vicinity of the resonances (a) $f_r$; (b) $f_r/2$; (c) $f_r/3$.}
    \label{ris:Chi3}
    \end{figure}

    \begin{figure}[t]
    \includegraphics[width=0.9\linewidth]{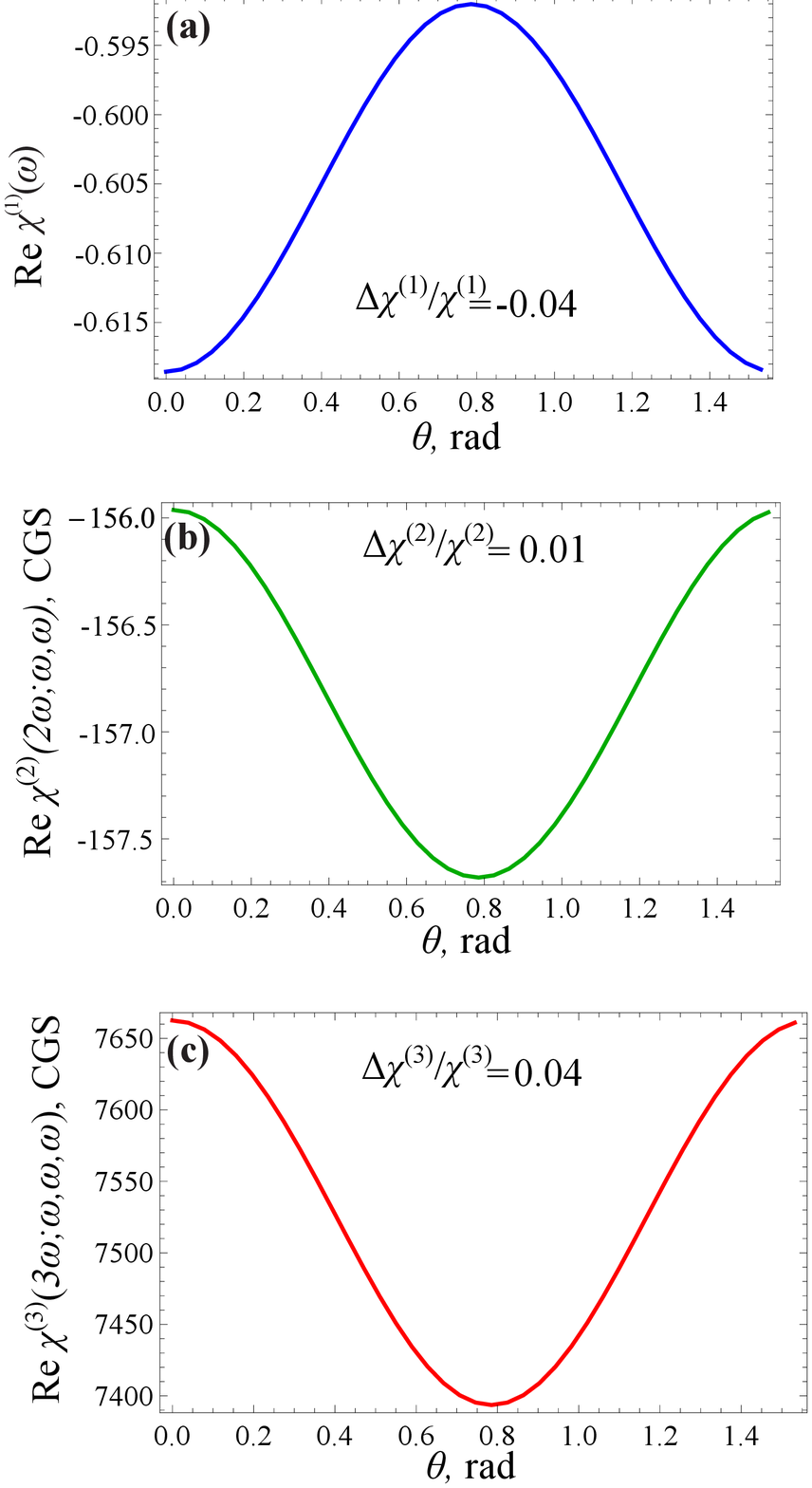}
    \caption{(Color online) Dependence of the real part of linear and nonlinear susceptibilities on the angle $\theta$ between the wave vector of the fundamental harmonic and $\Gamma X$ direction. Here, $f=f_r=0.616$~GHz, $\lambda/a=48$.}
    \label{ris:Chi-Angle}
    \end{figure}

Now we proceed to the demonstration of the main features of the developed theoretical model for a particular example. To this end we consider a 3D structure with the period $a=1$~cm composed of short wires (much shorter than the wavelength) loaded by varactors Skyworks SMV 1231-079~\refPR{Skyworks} possessing nonlinear capacitance as well as associated linear parameters. Varactor is inserted in the gap of the size $\Delta l=1.3$~mm in the middle of the wire with the radius $r=1$~mm and half-length $l=3$~mm. The total inductance of varactor inclusion is $L_{t}=42.5$~nH and the  capacitance determining the input impedance of the wire (without varactor capacitance) is $C_t=0.2$~pF. Other varactor parameters are set to those specified by the manufacturer~\refPR{Skyworks}. For clarity, dispersion diagram is shown for zero dissipation. However, nonlinear susceptibilities are calculated with realistic losses according to specifications~\refPR{Skyworks}.

The dispersion diagram of the described metamaterial in linear regime of operation calculated with Eq.~\eqref{DispEq} is presented in Fig.~\ref{ris:LinearDisp}. Analysis of the calculated diagram reveals that there are two frequency intervals where spatial dispersion effects are most pronounced. These frequency intervals correspond to the so-called mixed dispersion regime studied in detail for linear uniaxial metamaterials in Ref.~\refPR{Gorlach2014}. Mixed dispersion regime arises at frequencies in the vicinity of zeros~\refPR{Gorkunov-98,Belov2005} (low-$\varepsilon$ mixed regime) and poles~\refPR{landau8} (high-$\varepsilon$ mixed regime) of the structure local permittivity Eq.~\eqref{Clausius}.

We expect the most significant deviations of nonlinear susceptibilities from the local effective medium model to occur in mixed dispersion regime. Note that in the low-$\varepsilon$ mixed regime the main manifestation of spatial dispersion is the emergence of longitudinal waves propagating closely to $\Gamma$Z direction~\refPR{Gorlach2014}. Excitation of such longitudinal modes is not simple in experiment. Therefore, we concentrate on the analysis of transverse modes which are strongly affected by nonlocality in high-$\varepsilon$ mixed regime. In this example, high- and low-$\varepsilon$ mixed regimes correspond to the spectral ranges $0.565<f<0.670$~GHz and $0.992<f<1.024$~GHz, respectively, and frequency of linear permittivity resonance is $f_r=0.616$~GHz.

Perturbative analysis of the nonlinear oscillator model~\refPR{Boyd} suggests that nonlinear susceptibilities $\chi^{(2)}$ and $\chi^{(3)}$ are subject to resonant enhancement not only at the frequency $f_r$ but also at frequencies $f_r/2$ ($\chi^{(2)}$, $\chi^{(3)}$) and $f_r/3$ ($\chi^{(3)}$ only) as explained in more details in  Appendix. Therefore, we also studied metamaterial nonlinear properties in spectral intervals around $f_r/2$ and $f_r/3$. Linear and nonlinear susceptibilities calculated for $\Gamma$M direction of propagation by Eqs.~\eqref{Chi1}, \eqref{Chi2}, \eqref{Chi3} are compared with the predictions of the local effective medium model Eqs.~\eqref{Clausius}, \eqref{Chi2Loc}, \eqref{Chi3Loc} in Figs.~\ref{ris:Chi}, \ref{ris:Chi3}. The results show that significant deviations from the local effective medium model indeed occur in high-$\varepsilon$ mixed regime whereas at lower frequency resonances $f_r/2$ and $f_r/3$ resonant enhancement of nonlinear susceptibilities is accurately captured by the local model. Furthermore, it can be noticed that spatial dispersion dumps the resonance that leads to the decrease of the maximal achievable values of nonlinear susceptibility. Therefore, we conclude that spatial dispersion should be necessarily taken into account while describing nonlinearities of metamaterials in the vicinity of permittivity resonance.

Another interesting manifestation of spatial dispersion is the dependence of nonlinear susceptibilities on the direction of propagation of the fundamental wave with respect to the sample crystallographic axes. In Fig.~\ref{ris:Chi-Angle} we plot the dependence of linear and nonlinear susceptibilities on the angle between the wave vector $\vec{k}$ and $\Gamma$X direction in the first Brillouin zone of the crystal. Even though a dependence of nonlinear susceptibilities on the propagation direction is also known for photonic crystals~\refPR{Sakoda}, it should be stressed that the metamaterial operates in deeply subwavelength regime with the ratio $\lambda/a\approx 48$ at frequency $f_r$. Nevertheless, variation of susceptibility with the direction of wave vector $\vec{k}$ reaches $4\%$ for $\chi^{(1)}$ and $\chi^{(3)}$ and $1\%$ for $\chi^{(2)}$ at the resonance frequency $f_{\rm{r}}=0.616$~GHz. In general, at frequencies around $f_r$ maximal variation of susceptibilities $\chi^{(1)}$, $\chi^{(2)}$ and $\chi^{(3)}$ with the direction of wave vector is $4$-$5\%$.

Finally, as Fig.~\ref{ris:LinearDisp}b clearly shows, in high-$\varepsilon$ mixed regime there are two solutions of the dispersion equation corresponding to the given frequency and $\Gamma$M or $\Gamma$X directions of propagation. As a result, nonlinear susceptibilities Eqs.~\eqref{Chi2}, \eqref{Chi3} are multivalued functions of frequency in this spectral range. Consequently, one may expect that one TM-polarized incident beam can produce {\it two} second-harmonic (third-harmonic) beams with the same polarization and frequency. Importantly, the described  effect arises purely due to spatial dispersion effects and cannot be explained in the framework of the local effective medium model.

\section{Conclusions}\label{sec:Discussion}
We have developed a consistent theoretical approach for calculating effective nonlinear susceptibilities of nonlinear discrete metamaterials taking into account both frequency and spatial dispersion. We have modelled nonlinear metamaterial as a lattice of nonlinear uniaxial electric dipoles, and obtained closed-form expressions for effective nonlinear parameters. It has been demonstrated that spatial dispersion affects strongly nonlinear properties of metamaterials in the vicinity of effective permittivity resonance dumping the frequency  variation of nonlinear susceptibilities in comparison to the models of local effective media. We have predicted that, due to the spatial dispersion effects, one incident light beam may produce two harmonic beams with the same polarization. Additionally, we have demonstrated the dependence of nonlinear susceptibilities on the direction in which the fundamental harmonic propagates with respect to the crystallographic axes. Our results suggest that  nonlocality is important  in metamaterials even if they operate in a deeply subwavelength regime. Furthermore, we have verified that our results yield an accurate form of the so-called local field corrections to the nonlinear susceptibilities when spatial dispersion effects become negligible. Our conclusions are also valid for three-dimensional arrays of uniaxial magnetic scatterers such as varactor-loaded split-ring resonantors. We believe that our study provides important insights into the characterization of nonlinear metamaterials exhibiting large resonant nonlinearities.

\section{Acknowledgments}
This work was supported by the Government of the Russian Federation (Grant No.~074-U01), the Dynasty Foundation, a grant of the President of the Russian Federation (MD-7841.2015.2), the Ministry of Education and Science of the Russian Federation (Projects No. 14.584.21.0009 10, GZ No.~2014/190, GZ No.~3.561.2014/K), Russian Foundation for Basic Research (Project No.~15-02-08957 A), and the Australian Research Council. M.G. acknowledges a visiting appointment at the University of Technology Sydney.

\section*{Appendix. Linear and nonlinear polarizabilities of a short varactor-loaded wire}

    \begin{figure}[b]
    \includegraphics[width=0.7\linewidth]{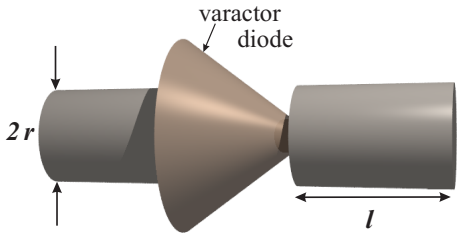}
    \caption{(Color online) A schematic representation of short varactor-loaded wire used as a building block of metamaterial.}
    \label{ris:Varactor}
    \end{figure}

In this appendix we derive expressions characterizing linear and nonlinear properties of a short ($2\,l\ll \lambda$) varactor-loaded wire. Varactor is described by the nonlinear capacitance as well as associated linear parameters. Nonlinearity in such meta-atom arises due to the voltage-dependent varactor capacitance that is well approximated by the formula~\refPR{Skyworks}:
\begin{equation}\label{CU-approximation}
C(U)=\frac{C_{J0}}{(1+U/U_J)^M}
\end{equation}
where $U$ is deemed positive in the case of reverse varactor bias. In the case of varactor SMV 1231-079, the parameters in Eq.~\eqref{CU-approximation} are as follows~\refPR{Skyworks}: $M=4.999$, $C_{J0}=1.88$~pF and $U_J=10.13$~V.

Making use of the definition $C=dq/dU$ one can relate reverse voltage on the varactor $U_V$ to its charge $q$:
\begin{equation}\label{UNonlinear}
U_V(q)=U_J\,\left[\left(1+\frac{1-M}{C_{J0}\,U_J}\,q\right)^{1/(1-M)}-1\right]\:.
\end{equation}
Under the assumption $|q|\ll C_{J0}\,U_J$ the right-hand side of Eq.~\eqref{UNonlinear} can be expanded in a series with respect to $q$:
\begin{equation}\label{UNonlinearAppr}
U_V(q)=\frac{1}{C_{J0}}\,\left[q+\frac{M\,q^2}{2\,C_{J0}\,U_J}+\frac{M\,(2M-1)}{6\,C_{J0}^2\,U_J^2}\,q^3\right]\:.
\end{equation}
Taking into account parasitic linear capacitance $C_p$ loaded parallel to varactor, Eq.~\eqref{UNonlinearAppr} can be further rearranged to yield
\begin{equation}\label{UNonlinearAppr2}
\begin{split}
& U_V(Q)=\frac{1}{C_{0}}\,\left[Q+\frac{M\,C_{J0}\,Q^2}{2\,C_{0}^2\,U_J}+\right.\\
& \left.\frac{C_{J0}\,Q^3}{C_0^4\,U_J^2}\,\left(C_0\,\frac{M(2M-1)}{6}-C_p\,\frac{M^2}{2}\right)\right]
\end{split}
\end{equation}
where $Q$ is the total charge stored by both varactor and parasitic capacitance $C_p$, and $C_0=C_{J0}+C_p$. Denoting the inductance and resistance of the entire varactor inclusion by $L_t$ and $R_t$, respectively, we obtain the nonlinear oscillator equation for the total charge stored in the system:
\begin{equation}\label{NonlinOscEq}
\ddot{Q}+2\,\beta_0\,\dot{Q}+\omega_0^2\,Q+\beta_2\,Q^2+\beta_3\,Q^3=\mathfrak{E}(t)
\end{equation}
where $\mathfrak{E}(t)=\mathfrak{\epsilon}(t)/L_t$, $\mathfrak{\epsilon}(t)$ is electromotive force, $\beta_0=R_t/(2\,L_t)$, $\omega_0=1/\sqrt{L_t\,C_0}$,
\begin{equation}\label{Beta2}
\beta_2=\frac{\omega_0^2\,M\,C_{J0}}{2\,C_{0}^2\,U_J}
\end{equation}
and
\begin{equation}\label{Beta3}
\beta_3=\frac{\omega_0^2\,C_{J0}}{C_0^4\,U_J^2}\,\left[C_0\,\frac{M(2M-1)}{6}-C_p\,\frac{M^2}{2}\right]\:.
\end{equation}

To calculate linear and nonlinear polarizabilities, we consider the system composed of nonlinear varactor with associated linear parameters and wires. Denoting the input impedance of the entire wire by $Z_{\rm{inp}}(\omega)=-1/(i\omega\,C_t)$ we obtain

\begin{equation}\label{NonlinOscEq2}
\begin{split}
& \ddot{Q}+2\,\beta_0\,\dot{Q}+\omega_0^2\,Q+\beta_2\,Q^2+\beta_3\,Q^3=\\
& \text{Re}\left\lbrace\frac{\xi\,e^{-i\omega\,t}}{L_t}\,\left[l\,E(\omega)-I_0(\omega)\,Z_{\rm{inp}}(\omega)\right]+\right.\\
& \left.\frac{\xi^2\,e^{-2i\,\omega t}}{L_t}\,\left[l\,E(2\omega)-I_0(2\omega)\,Z_{\rm{inp}}(2\,\omega)\right]+\right.\\
& \left.\frac{\xi^3\,e^{-3i\,\omega t}}{L_t}\,\left[l\,E(3\omega)-I_0(3\omega)\,Z_{\rm{inp}}(3\,\omega)\right]\right\rbrace-\\
&-\frac{\xi^2\,U_l(0)}{L_t}\:,
\end{split}
\end{equation}
where $I_0(\Omega)=\dot{Q}_\Omega$, $U_l(0)=\lim\limits_{\omega\rightarrow 0} I_0(\omega)\,Z_{\rm{inp}}(\omega)$ is a static voltage arising on varactor inclusion and $\xi$ is an auxiliary dimensionless parameter that is usually introduced for the perturbative solution of nonlinear oscillator equation and which will be set to $1$ at the end of the calculation~\refPR{Boyd}. We consider the incident field $E(\omega)$ as sufficiently small. In this case the steady-state solution for Eq.~\eqref{NonlinOscEq2} may be searched in the form of a power series in $\xi$
\begin{equation}\label{Ansatz}
Q(t)=\xi\,Q^{(1)}(t)+\xi^2\,Q^{(2)}(t)+\xi^3\,Q^{(3)}(t)\:.
\end{equation}
Plugging ansatz Eq.~\eqref{Ansatz} into Eq.~\eqref{NonlinOscEq2} yields the set of equations:
\begin{equation}\label{SetNonlinEq1}
\begin{split}
& \ddot{Q}^{(1)}+2\,\beta_0\,\dot{Q}^{(1)}+\omega_0^2\,Q^{(1)}=\\
& \text{Re}\left\lbrace\frac{e^{-i\omega\,t}}{L_t}\,\left[l\,E(\omega)-I_0'(\omega)\,Z_{\rm{inp}}(\omega)\right]\right\rbrace\:,
\end{split}
\end{equation}
\begin{equation}\label{SetNonlinEq2}
\begin{split}
& \ddot{Q}^{(2)}+2\,\beta_0\,\dot{Q}^{(2)}+\omega_0^2\,Q^{(2)}+\beta_2\,\left[Q^{(1)}\right]^2=\\
& \text{Re}\left\lbrace\frac{e^{-2i\omega\,t}}{L_t}\,\left[l\,E(2\,\omega)-I_0(2\,\omega)\,Z_{\rm{inp}}(2\,\omega)\right]\right\rbrace-\frac{U_l(0)}{L_t}\:,
\end{split}
\end{equation}
\begin{equation}\label{SetNonlinEq3}
\begin{split}
& \ddot{Q}^{(3)}+2\,\beta_0\,\dot{Q}^{(3)}+\omega_0^2\,Q^{(3)}+2\,\beta_2\,Q^{(1)}\,Q^{(2)}+\\
& \beta_3\,\left[Q^{(1)}\right]^3=\text{Re}\left\lbrace\frac{e^{-3i\omega\,t}}{L_t}\,\left[l\,E(3\,\omega)-I_0(3\,\omega)\,Z_{\rm{inp}}(3\,\omega)\right]-\right.\\
&\left.\frac{e^{-i\omega\,t}}{L_t}\,I_0''(\omega)\,Z_{\rm{inp}}(\omega)\right\rbrace
\end{split}
\end{equation}
where $I_0'(\omega)=-i\omega\,Q^{(1)}(\omega)$ and $I_0''(\omega)=-i\omega\,Q^{(3)}(\omega)$, $I_0'(\omega)+I_0''(\omega)=I_0(\omega)$. Each of Eqs.~\eqref{SetNonlinEq1}, \eqref{SetNonlinEq2}, \eqref{SetNonlinEq3} is a linear differential equation with single unknown function ($Q^{(1)}(t)$, $Q^{(2)}(t)$ and $Q^{(3)}(t)$, respectively). The solutions of these equations are as follows:
\begin{gather}
Q^{(1)}(t)=\text{Re}\left(x_1\,e^{-i\omega\,t}\right)\:,\label{SetSolutions1}\\
Q^{(2)}(t)=x_0+\text{Re}\left(x_2\,e^{-2i\omega\,t}\right)\:,\label{SetSolutions2}\\
Q^{(3)}(t)=\text{Re}\left(x_1'\,e^{-i\omega\,t}+x_3\,e^{-3i\omega\,t}\right)\label{SetSolutions3}
\end{gather}
with the amplitudes $x$ defined as
\begin{gather}
x_1=\frac{l\,E(\omega)}{F(\omega)}\:,\label{x1}\\
x_0=-\frac{\beta_2\,l^2\,L_t\,\left|E(\omega)\right|^2}{2\,F(0)\,\left|F(\omega)\right|^2}\:,\label{x0}\\
x_2=\frac{l\,E(2\,\omega)}{F(2\,\omega)}-\frac{\beta_2\,l^2\,L_t\,E^2(\omega)}{2\,F^2(\omega)\,F(2\,\omega)}\:,\label{x2}\\
x_1'=-\frac{\beta_2\,l^2\,L_t}{\left|F(\omega)\right|^2\,F(2\,\omega)}\,E^*(\omega)\,E(2\,\omega)+\notag\\
\frac{l^3\,\left|E(\omega)\right|^2\,E(\omega)}{\left|F(\omega)\right|^2\,F^2(\omega)}\,\left[-\frac{3\,\beta_3\,L_t}{4}+\frac{\beta_2^2\,L_t^2}{F(0)}+\frac{\beta_2^2\,L_t^2}{2\,F(2\,\omega)}\right]\:,\label{x12}\\
x_3=\frac{l\,E(3\,\omega)}{F(3\,\omega)}-\frac{\beta_2\,l^2\,L_t\,E(\omega)\,E(2\,\omega)}{F(\omega)\,F(2\,\omega)\,F(3\,\omega)}+\notag\\
\left[\frac{\beta_2^2\,l^3\,L_t^2}{2\,F^3(\omega)\,F(2\,\omega)\,F(3\,\omega)}-\frac{\beta_3\,l^3\,L_t}{4\,F^3(\omega)\,F(3\,\omega)}\right]\,E^3(\omega)\:.\label{x3}
\end{gather}

where we use the designation
\begin{gather}
F(\Omega)=L_t\,D(\Omega)-i\Omega\,Z_{\rm{inp}}(\Omega)=L_t\,D(\Omega)+\frac{1}{C_t}\:,\\
D(\Omega)=\omega_0^2-2i\,\beta_0\,\Omega-\Omega^2\:.
\end{gather}
Having solved the nonlinear oscillator equation and assuming the current distribution as in a short symmetric antenna~\refPR{Antenna}, we are able to calculate the meta-atom dipole moment as $d(\omega)=l\,\left(x_1+x_1'\right)$, $d(2\,\omega)=l\,x_2$ and $d(3\,\omega)=l\,x_3$. It is now straightforward to prove the validity of Eqs.~\eqref{d1}, \eqref{d2}, \eqref{d3} used in Sec.~\ref{sec:Homogenization} for the meta-atom characterization. Linear and nonlinear meta-atom polarizabilities are given by the formulas:

\begin{gather}
\alpha_1(\omega)=\frac{l^2}{F(\omega)}\:,\label{Al1}\\
\alpha_2(\omega;2\,\omega,-\omega)=-\frac{1}{2}\,\frac{\beta_2\,l^3\,L_t}{\left|F(\omega)\right|^2\,F(2\,\omega)}\:,\label{Al21}\\
\alpha_2(2\,\omega;\omega,\omega)=-\frac{\beta_2\,l^3\,L_t}{2\,F^2(\omega)\,F(2\,\omega)}\:,\label{Al2}\\
\alpha_2(3\,\omega;2\,\omega,\omega)=-\frac{\beta_2\,l^3\,L_t}{2\,F(3\,\omega)\,F(2\,\omega)\,F(\omega)}\:,\label{Al23}\\
\alpha_3(\omega;\omega,\omega,-\omega)=\frac{l^4}{\left|F(\omega)\right|^2\,F^2(\omega)}\times\notag\\
\left[-\frac{\beta_3\,L_t}{4}+\frac{\beta_2^2\,L_t^2}{3\,F(0)}+\frac{\beta_2^2\,L_t^2}{6\,F(2\,\omega)}\right]\:,\label{Al31}\\
\alpha_3(3\,\omega;\omega,\omega,\omega)=\frac{l^4}{F(3\,\omega)\,F^3(\omega)}\,\left[-\frac{\beta_3\,L_t}{4}+\frac{\beta_2^2\,L_t^2}{2\,F(2\,\omega)}\right]\:.
\end{gather}

The derived expressions suggest that nonlinear polarizabilities exhibit resonant enhancement at frequencies satisfying one of the following conditions $F(\omega)=0$ or  $F(2\,\omega)=0$ or $F(3\,\omega)=0$ that was used in Sec.~\ref{sec:Numerical}.

\bibliography{NonlinearLib}

\end{document}